# Optimizing decision making for soccer line-up by a quantum annealer

*Dr. Aitzol Iturrospe (aiturrospe@spdtek.com)*

## Abstract

In this paper an application of D-Wave Systems' quantum annealer for optimizing initial line-up of a soccer team is presented. Players and their playing position are selected to maximize the sum of players ratings for both a 4-3-3 attack formation and a 4-2-3-1 medium defensive formation. The problem is stated as a binary quadratic model (BQM) and it is solved in a D-Wave Leap's Hybrid Solver. Results are presented and compared with already published results obtained with a classical solver.

## 1. Introduction

Coach line-up decision making is of utmost importance for the performance of sport teams. One important decision that must be taken by a soccer coach is to determine the starting line-up of players. The coach considers many factors of each player and the team strategy, so it is a complex decision-making process (Saaty, 1994). The team formation describes how the players are positioned on the soccer field. Some players can play in more than one position, even if their valuation can be different for each position. The decision of the coach can impact team performance reducing their chances to win the match (Purwanto et al., 2018), e.g. if the appropriate players are not selected for a given formation or they are placed in positions where they cannot give their best.

In the article (Mahrudinda et al., 2020), a line-up optimization method is proposed, and it is applied to optimize Liverpool FC players selection for the initial line-up during the 2020/2021 Premier League. The authors used binary integer programming (BIP) modeling, representing if a player is selected to play in a specific position as a binary variable. Two different formations are considered. Firstly, a 4-3-3 formation with one goalkeeper, four defenders, three midfielders and three forwards. Secondly, a 4-3-2-1 formation with one goalkeeper, four defenders, three midfielders, two attack midfielders and a forward striker.

Quantum annealing processors (QAP) are suited to be used in optimization problems, as long as the objective function is minimized over binary variables, or a set of discrete variables (DWAVE, 2021). The objective function is stated as a Hamiltonian, a mathematical description of some physical system in terms of its energies, and the QAP seek a solution that minimizes the energy of the Hamiltonian. For most non-convex Hamiltonians, finding the minimum energy state is an NP-hard problem that classical computers cannot solve efficiently. By defining the optimization problem as a quadratic unconstrained binary optimization (QUBO), it can be mapped to the qubits and couplers in the QAP. BQM stands for Binary Quadratic Model and is just a general term that encompasses both Ising and QUBO problems. Leap's quantum-classical hybrid BQM solvers are intended to solve arbitrary application problems formulated as BQM. These solvers accept arbitrarily structured, unconstrained problems formulated as BQMs, with any constraints typically represented through penalty models.

Binary optimization problems such as the stated by (Mahrudinda et al., 2020) are suitable to be reformulated as BQM and solved in quantum-classical hybrid BQM solver. The problem is restated in section 2 in an appropriate way to be solved in a QAM. In section 3 the optimization problem is reformulated as a BQM in order to obtain a Hamiltonian that can be solved by the quantum annealer. In section 4, the problem is solved by a D-Wave Leap's Hybrid solver, discussing the optimization results, and comparing them to those presented in (Mahrudinda et al., 2020). Finally, in the last section some general conclusions are stated, and likely future research lines are proposed.

## 2. Problem statement

Soccer is a game played by two teams where each team consists of eleven players. The eleven players chosen to play at the start of the game are called starting lineups.

The eleven soccer players are divided into several positions in accordance with the team formation. In addition to one goalkeeper (GK), the players are divided into three main positions, defenders (D), midfielders (M), and forward/strikers (FW). Each major position can be subdivided into several more specific positions, e.g. for the defender is divided into central defender (DC), left wing defender (DL) and right wing defender (DR). The midfielder's position can be divided into defensive midfielder (DM), central midfielder (CM) and attack midfielder (AM). The forward's positions might be divided into right wing forward (FWR), left wing forward (FWL) and forward (FW).

Figure 1 shows players' ratings depending on the position presented in the article (Mahrudinda et al., 2020). There are 43 ratings for (player, position) pairs. Each pair (player, position) will be considered as a binary variable; being 1 if the player is lined up to play in that position and 0 otherwise. A vector of binary variables $x_i$ will be considered, starting from the upper left of the table, and going through the table column wise first and then row wise.

| Player | Position | | | | | | | | | |
|---|---|---|---|---|---|---|---|---|---|---|
| | GK | DC | DL | DR | DM | CM | AM | FWL | FWR | FW |
| Alisson | 6.81 | - | - | - | - | - | - | - | - | - |
| Adrián | 5.86 | - | - | - | - | - | - | - | - | - |
| van Dijk | - | 6.62 | - | - | - | - | - | - | - | - |
| Gomez | - | 6.91 | - | - | - | - | - | - | - | - |
| Robertson | - | - | 6.85 | - | 6.97 | 6.64 | - | - | - | - |
| Matip | - | 6.69 | - | - | - | - | - | - | - | - |
| Phillips | - | 7.24 | - | - | - | - | - | - | - | - |
| Arnold | - | - | - | 6.93 | 6.4 | - | - | - | - | - |
| Williams | - | - | - | 6.63 | 7.66 | 7.77 | - | - | - | - |
| Fabinho | - | 7.11 | - | - | - | 6.58 | - | - | - | - |
| Wijnaldum | - | - | - | - | 6.64 | 6.74 | 7.16 | - | - | - |
| Thiago | - | - | - | - | - | 7.38 | - | - | - | - |
| Milner | - | - | - | 8.15 | - | 6.97 | - | - | - | - |
| Keïta | - | - | - | - | - | 6.43 | - | - | - | - |
| Henderson | - | - | - | - | 6.8 | 6.7 | - | - | - | - |
| Jones | - | - | - | - | - | 7.13 | - | - | - | - |
| Shaqiri | - | - | - | - | - | 6.49 | 6.52 | - | - | - |
| Firmino | - | - | - | - | - | - | 6.8 | - | - | 6.99 |
| Mané | - | - | - | - | - | - | 7.24 | 7.56 | - | - |
| Salah | - | - | - | - | - | - | - | - | 7.42 | 6.8 |
| Minamino | - | - | - | - | - | 6.05 | - | - | - | 6.03 |
| Jota | - | - | - | - | - | 9.39 | 6.77 | 7.23 | 7.84 | 8.22 |
| Origi | - | - | - | - | - | - | - | 6.4 | - | 5.84 |

*Figure 1. Players' ratings depending on the position they play (Mahrudinda et al., 2020).*

The objective function to be maximized is the total sum of ratings of selected players, subject to the constraints defined for both formations considered (see table 1 and table 2).

$$\text{Maximize } H_Z = 6.81x_1 + 5.86x_2 + 6.62x_3 + \cdots + 6.03x_{41} + 8.22x_{42} + 5.84x_{43} \quad (1)$$

*Table 1. Constraints for a formation 4-3-3*

| | |
|---|---|
| $x_1 + x_2 + x_3 + \cdots + x_{41} + x_{42} + x_{43} = 11$ | 11 players |
| $x_1 + x_2 = 1$ | 1 goalkeeper |
| $x_3 + x_4 + x_5 + x_6 + x_7 = 2$ | 2 central defenders |
| $x_8 = 1$ | 1 left-hand side defender |
| $x_9 + x_{10} + x_{11} = 1$ | 1 right-hand side defender |
| $x_{17} + x_{18} + x_{19} + \cdots + x_{26} + x_{27} + x_{28} = 3$ | 3 central midfielders |
| $x_{34} + x_{35} + x_{36} = 1$ | 1 left forward |
| $x_{37} + x_{38} = 1$ | 1 right forward |
| $x_{39} + x_{40} + x_{41} + x_{42} + x_{43} = 1$ | 1 forward/striker |

*Table 2. Constraints for a formation 4-3-2-1*

| | |
|---|---|
| $x_1 + x_2 + x_3 + \cdots + x_{41} + x_{42} + x_{43} = 11$ | 11 players |
| $x_1 + x_2 = 1$ | 1 goalkeeper |
| $x_3 + x_4 + x_5 + x_6 + x_7 = 2$ | 2 central defenders |
| $x_8 = 1$ | 1 left-hand side defender |
| $x_9 + x_{10} + x_{11} = 1$ | 1 right-hand side defender |
| $x_{12} + x_{13} + x_{14} + x_{15} + x_{16} = 2$ | 2 defensive midfielders |
| $x_{29} + x_{30} + x_{31} + x_{32} + x_{33} = 3$ | 3 attacking midfielder |
| $x_{39} + x_{40} + x_{41} + x_{42} + x_{43} = 1$ | 1 forward/striker |

For both formations, we impose the following inequalities as constraints to avoid QPU solutions with players in multiple positions.

*Table 3. Constraints to avoid solutions where the same player plays in different positions*

| | |
|---|---|
| $x_8 + x_{12} + x_{17} \leq 1$ | (I1) |
| $x_9 + x_{13} \leq 1$ | (I2) |
| $x_{10} + x_{14} + x_{18} \leq 1$ | (I3) |
| $x_{15} + x_{20} + x_{29} \leq 1$ | (I4) |
| $x_{11} + x_{22} \leq 1$ | (I5) |
| $x_{16} + x_{24} \leq 1$ | (I6) |
| $x_{26} + x_{30} \leq 1$ | (I7) |
| $x_{32} + x_{34} \leq 1$ | (I8) |
| $x_{37} + x_{40} \leq 1$ | (I9) |
| $x_{28} + x_{33} + x_{35} + x_{38} + x_{42} \leq 1$ | (I10) |

### 3. BQM formulation

As the objective function is a maximization function, it is converted to a minimization by multiplying the entire expression by -1.

$$\arg\ min(-H_Z) \quad (2)$$

Equality constraints $\sum_{i=0}^{N} a_i x_i = b$ are standardly formulated in BQMs as minimizing $\left(\sum_{i=0}^{N} a_i x_i - b\right)^2$ functions. Therefore, the constraints in table 1 and 2 are reformulated following then quadratic formulation and they are shown in tables 4, 5 and 6.

*Table 4. Reformulated constraints as functions which are common for both formations*

| | | |
|---|---|---|
| $(x_1 + x_2 + x_3 + \cdots + x_{41} + x_{42} + x_{43} - 11)^2$ | 11 players | $(C_1)$ |
| $(x_1 + x_2 - 1)^2$ | 1 goalkeeper | $(C_2)$ |
| $(x_3 + x_4 + x_5 + x_6 + x_7 - 2)^2$ | 2 central defenders | $(C_3)$ |
| $(x_8 - 1)^2$ | 1 left-hand side defender | $(C_4)$ |
| $(x_9 + x_{10} + x_{11} - 1)^2$ | 1 right-hand side defender | $(C_5)$ |
| $(x_{39} + x_{40} + x_{41} + x_{42} + x_{43} - 1)^2$ | 1 forward/striker | $(C_6)$ |

*Table 5. Reformulated constraints as functions for a formation 4-3-3*

| | | |
|---|---|---|
| $(x_{17} + x_{18} + x_{19} + \cdots + x_{26} + x_{27} + x_{28} - 3)^2$ | 3 central midfielders | $(C_7)$ |
| $(x_{34} + x_{35} + x_{36} - 1)^2$ | 1 left forward | $(C_8)$ |
| $(x_{37} + x_{38} - 1)^2$ | 1 right forward | $(C_9)$ |

*Table 6. Reformulated constraints as functions for a formation 4-3-2-1*

| | | |
|---|---|---|
| $(x_{12} + x_{13} + x_{14} + x_{15} + x_{16} - 2)^2$ | 2 defensive midfielders | $(C_{10})$ |
| $(x_{29} + x_{30} + x_{31} + x_{32} + x_{33} - 3)^2$ | 3 attacking midfielder | $(C_{11})$ |

For inequality constraints, slack variables are introduced in order to reduce them to equalities (DWAVE, 2021) as follows:

*Table 7. Inequality constraints reformulated as equalities for the BQM.*

| | |
|---|---|
| $(x_8 + x_{12} + x_{17} + a_1 - 1)^2$ | $(I_1)$ |
| $(x_9 + x_{13} + a_2 - 1)^2$ | $(I_2)$ |
| $(x_{10} + x_{14} + x_{18} + a_3 - 1)^2$ | $(I_3)$ |
| $(x_{15} + x_{20} + x_{29} + a_4 - 1)^2$ | $(I_4)$ |
| $(x_{11} + x_{22} + a_5 - 1)^2$ | $(I_5)$ |
| $(x_{16} + x_{24} + a_6 - 1)^2$ | $(I_6)$ |
| $(x_{26} + x_{30} + a_7 - 1)^2$ | $(I_7)$ |
| $(x_{32} + x_{34} + a_8 - 1)^2$ | $(I_8)$ |
| $(x_{37} + x_{40} + a_9 - 1)^2$ | $(I_9)$ |
| $(x_{28} + x_{33} + x_{35} + x_{38} + x_{42} - 1)^2$ | $(I_{10})$ |

The Lagrange multiplier ($\lambda_i$) acts as a weight given to the constraint. It should be set high enough to ensure the constraint is satisfied but setting it too high obscures the real function we are trying to minimize. All the Lagrange multipliers were set equal to each other and further equal to eleven times the maximum rating, thus 90.5.

The statement for the BQM objective function for formation 4-3-3 was $H_{433} = -H_Z + \lambda\left(\sum_{i=1}^{9} C_i + \sum_{i=1}^{9} I_i\right)$ and $H_{4321} = -H_Z + \lambda\left(\sum_{i=1}^{6} C_i + \sum_{i=10}^{11} C_i + \sum_{i=1}^{9} I_i\right)$ for a 4-3-2-1 formation.

## 4. Results

Both Hamiltonians, $H_{433}$ and $H_{4321}$, were sampled in a Hybrid BQM solver (e.g., hybrid_binary_quadratic_model_version2), being QPU access time 1.96 s and 1.93 s respectively.

All imposed constraints, both linear equations and inequalities, were fulfilled. Therefore, the minimized energy were $H_{433} = -H_Z = -82.67$ and $H_{4321} = -H_Z = -80.04$.

In table 8 and 9 the best sampled line-ups are presented for formation 4-3-3 and 4-3-2-1 respectively. Both line-ups are compared with the optimization results presented in (Mahrudinda et al., 2020). It can be observed that though line-ups coincide for formation 4-3-2-1, a small improvement was obtained by the quantum annealer optimization for formation 4-3-3; by replacing Gomez for Fabinho as a central defender.

*Table 8. Line-up obtained from QPU optimization compared to optimized line-up presented in (Mahrudinda et al., 2020) for formation 4-3-3*

| QPU results for 4-3-3 formation | | | | Results for 4-3-3 formation in (Mahrudinda et al., 2020) | | | |
|---|---|---|---|---|---|---|---|
| **Binary variable** | **Player Name** | **Position** | *Rating* | **Binary variable** | **Player Name** | **Position** | *Rating* |
| $x1$ | Alisson | GK | 6.81 | $x1$ | Alisson | GK | 6.81 |
| $x6$ | Philips | DC | 7.24 | $x4$ | Gomez | DC | 6.91 |
| $x7$ | Fabinho | DC | 7.11 | $x6$ | Philips | DC | 7.24 |
| $x8$ | Robertson | DL | 6.85 | $x8$ | Robertson | DL | 6.85 |
| $x11$ | Milner | DR | 8.15 | $x11$ | Milner | DR | 8.15 |
| $x18$ | Williams | CM | 7.77 | $x18$ | Williams | CM | 7.77 |
| $x21$ | Thiago | CM | 7.38 | $x21$ | Thiago | CM | 7.38 |
| $x28$ | Jota | CM | 9.39 | $x28$ | Jota | CM | 9.39 |
| $x34$ | Mane | FWL | 7.56 | $x34$ | Mane | FWL | 7.56 |
| $x37$ | Salah | FWR | 7.42 | $x37$ | Salah | FWR | 7.42 |
| $x39$ | Firminho | FW | 6.99 | $x39$ | Firminho | FW | 6.99 |
| **Max $H_Z$** | | | **82.67** | **Max $H_Z$** | | | **82.47** |

*Table 9. Line-up obtained from QPU optimization compared to optimized line-up presented in (Mahrudinda et al., 2020) for formation 4-3-2-1*

| QPU results for 4-3-2-1 formation | | | | Results for 4-2-3-1 formation in (Mahrudinda et al., 2020) | | | |
|---|---|---|---|---|---|---|---|
| **Binary variable** | **Player Name** | **Position** | *Rating* | **Binary variable** | **Player Name** | **Position** | *Rating* |
| $x1$ | Alisson | GK | 6.81 | $x1$ | Alisson | GK | 6.81 |
| $x6$ | Philips | DC | 7.24 | $x6$ | Philips | DC | 7.24 |
| $x7$ | Fabinho | DC | 7.11 | $x7$ | Fabinho | DC | 7.11 |
| $x8$ | Robertson | DL | 6.85 | $x8$ | Robertson | DL | 6.85 |
| $x11$ | Milner | DR | 8.15 | $x11$ | Milner | DR | 8.15 |
| $x14$ | Williams | DM | 7.66 | $x14$ | Williams | DM | 7.66 |
| $x16$ | Henderson | DM | 6.80 | $x16$ | Henderson | DM | 6.80 |
| $x29$ | Wijnaldum | AM | 7.16 | $x29$ | Wijnaldum | AM | 7.16 |
| $x31$ | Firminho | AM | 6.80 | $x31$ | Firminho | AM | 6.80 |
| $x32$ | Mane | AM | 7.24 | $x32$ | Mane | AM | 7.24 |
| $x42$ | Jota | FW | 8.22 | $x42$ | Jota | FW | 8.22 |
| **Max Z** | | | **80.04** | **Max Z** | | | **80.04** |

## 5. Conclusions

D-Wave quantum annealers are suitable to solve optimization problems, in terms of objectives and constraints, as long as they can be formulated as BQM. The objective function must be minimized or maximized over a vector of binary variables mapped to qubits in the quantum computer.

As decision making about a line-up can be understood as selecting players to play in a given position or sidelining them, the optimization of a line-up seems suitable to be represented by a BQM.

The optimization problem presented in this paper could be extended to more complicated cases, e.g., considering opposing team's formations and player or as a tool to help team recruit new players.